\documentclass{kapproc} 

\usepackage{procps} 

\usepackage[dvips]{graphicx}

\upperandlowercase

\kluwerbib 

\begin{document}

\articletitle[TESIS - The TNG EROs Spectroscopic Identification Survey]
{TESIS - The TNG EROs Spectroscopic\\ Identification Survey}

\author{P. Saracco$^1$, M. Longhetti$^1$, R. Della Ceca$^1$, P. Severgnini$^1$, V. Braito$^1$, R. Bender$^2$, N. Drory$^3$, G. Feulner$^2$, U. Hopp$^2$, F. Mannucci$^4$, C. Maraston$^5$}
\affil{$^1$INAF - Osservatorio Astronomico di Brera, Via Brera 28, 20121 Milano, Italy\\ $^2$Universit\"ats-Sternwarte M\"unchen, Scheiner Str. 1, 81679 M\"unchen, Germany\\
$^3$University of Texas at Austin, Austin, Texas 78712\\
$^4$IRA-CNR, Largo E. Fermi 5, 50125 Firenze, Italy\\
 $^5$Max-Plank-Institut fuer  extraterrestrische Physik, Garching bei Munchen, Germany}


\anxx{Beerends\, John G.}



\section{The TESIS project}
The main scientific aims of the TESIS project are:\\
$i$) to measure the comoving density of massive
($\mathcal{M}_{star}>10^{11}M_\odot$) elliptical galaxies at $z>1$;\\
$ii$) to study the properties of X-ray emitting Extremely Red Objects (EROs);\\
$iii$) to measure the star formation rate (SFR) of dusty EROs probing
their connection with ULIRGs.\\
To this end we have started a near-IR very low resolution 
($\lambda/\Delta\lambda\simeq50$) spectroscopic survey of a complete sample 
of ~30 bright (K$<$18.5) EROs (Saracco et al. 2003, A\&A 398, 127).
The sample has been selected from the Munich Near-IR Cluster Survey 
(MUNICS, Drory et al. 2001, MNRAS 325, 550) over two fields 
($\sim$360 arcmin$^2$) covered by B, V, R, I, J and 
K-band observations.
The red optical-to-near-IR colors (R-K$\ge$5.3) allow to select z>1 evolved 
stellar systems; the bright K magnitudes assure the selection of 
massive galaxies and the near-IR spectra allow the detection of the 4000Å 
break at z>1.
The survey is carried out at the 3.6m Italian Telescopio Nazionale Galileo 
(TNG) and employes the prism disperser AMICI. 
It is designed to cover the full 
spectral range (8000-25000 \AA) in a single shot thus resulting extremely 
efficient in detecting continuum breaks and in describing spectral shapes.
Up to now, 40\% of the sample have been spectroscopically observed.

In parallel with the near-IR spectroscopic follow-up we obtained 150 ks of 
XMM-Newton observations for the two selected fields (75 ks each) to study the 
nature of the X-ray emitting EROs. 
The first set of XMM observations has been carried out in February 2003 while
the second set is expected in fall 2004.

\section[The density of $\mathcal{M}_{star}>10^{11}M_\odot$) ellipticals
at $z>1$]
{The density of $\mathcal{M}_{star}$$>$$10^{11}M_\odot$ ellipticals
at $z>1$}
We  classified 10 out of the 13 EROs observed so far: 7 early-type galaxies 
and 3 starbursts.
The properties of the 7 early-type galaxies are summarized in Tab. 1.
Assuming $\mathcal{M}/$L$_K$=0.5 [M/L]$_\odot$, all of the 7 galaxies 
have already formed and assembled a stellar mass well in excess to
$10^{11}M_\odot$.
Thus they would populate the very bright end 
(L$_{z=0}>$2L$^*$) of the local luminosity function of galaxies 
(we considered M$^*_K$=-24.2 from Cole et al. 2001,  
H$_0$=70 Km s$^{-1}$ Mpc$^{-1}$, $\Omega_m=0.3$ and $\Omega_\lambda=0.7$).
These 7 ellipticals account for a comoving density of about 
2.6$\times10^{-5}$ Mpc$^{-3}$. 
The number density of local L$>$2L$^*$ is 7$\times10^{-5}$ Mpc$^{-3}$
(we used $\phi^*_{E/S0}=1.5\times10^{-3}$ Mpc$^{-3}$, from Marzke et al. 1998).
Thus, they account for almost 40\% of the local population of massive 
elliptical galaxies (Longhetti et al. 2003, in prep.).
Since only 40\% of the sample of EROs has been observed we expect that
this density doubles at least. Hence, it is reasonable to expect that the
density of $\mathcal{M}_{star}$$>$$10^{11}M_\odot$ ellipticals at $z>1$
is consistent with the local one.
\begin{table}[ht]
\caption{Photometric and physical properties of the 7 early-type galaxies.}
\begin{center}
\begin{tabular}{lrccccc}
\hline
EROs-ID& R-K'& K'& \it z&M$_K$& $\mathcal{M}_{star}$&L$_K^{z=0}$\cr
\hline
       & mag & mag&     &mag  & 10$^{11}$M$_\odot$ & L$^*_K$\cr
\hline
S2F1\_142&6.0   &17.8&1.40&-26.2&3.5&$>$2.5\cr
S2F1\_357&6.0   &17.8&1.30&-26.0&3.0&$>$2.0\cr
S2F1\_389&$>$6.0&18.2&1.35&-25.7&2.0&$>$2.0\cr
S2F1\_527&5.9   &18.3&1.40&-25.7&2.0&$>$2.0\cr
S2F5\_109&5.3   &16.7&1.20&-26.9&6.5&$>$4.8\cr
S7F5\_45&5.8   &17.6&1.45&-26.4&4.3&$>$3.0\cr
S7F5\_254&$>$6.0&17.8&1.22&-25.8&2.5&$>$2.0\cr
\hline
\end{tabular}
\end{center}
\begin{tablenotes}
\end{tablenotes}
\end{table}
\section[Are type 2 QSO hidden in X-ray emitting EROs ?]
{Are type 2 QSOs hidden in X-ray emitting EROs ?}
The preliminary analysis of the 75 ks XMM observation centered on one
of the two selected fields shows that 5 EROs have a secure X-ray counterpart
down to a 2-10 keV flux limit of $\sim10^{15}$ erg cm$^{-2}$ s$^{-1}$.
None of them has been yet observed spectroscopically.
Their X-ray-to-optical flux ratios and their 2-10 keV luminosities 
 suggest the presence of an AGN in all of them.
For three out of the 5 EROs it has been possible to perform
a complete X-ray spectral analysis. 
The data are well fitted by a single power-law ($\Gamma\ge1.5$)
which provides  column densities N$_H\ge2\times10^{22}$ cm$^{-2}$ and 
intrinsic luminosities 
L$_{(2-10 keV)}\ge1.2\times10^{44}$ cgs.
These results indicate the presence of high luminosity, obscured AGNs, i.e.
QSO2 candidates (Severgnini et al. 2003, in prep.).

\end{document}